\documentstyle[12pt]{article}
\begin{document}
\noindent
{\Large\bf Quasireal photons in nucleus-nucleus collisions at intermediate
and high energies, applications to nuclear structure,  nuclear astrophysics
and particle physics} \\
{\large\em G.~Baur} \\
{\small Institut f\"{u}r Kernphysik, Forschungszentrum J\"{u}lich, Germany} \\
{\large\em S.~Typel, H.~H.~Wolter} \\
{\small Sektion Physik, Universit\"{a}t M\"{u}nchen, Garching, Germany} \\
\normalsize

\section{Introduction}

It is interesting to study nuclear collisions where the colliding nuclei
interact only electromagnetically. This can be achieved
by using bombarding energies below the Coulomb barrier or by choosing
very forward scattering angles in high energy collisions. With increasing
beam energy states at higher excitation energies can be excited; this can
lead, in addition to Coulomb excitation, also to Coulomb dissociation
\cite{Ber01}.
Such experiments are also feasible with secondary (radioactive) beams.
Because Coulomb dissociation is generally well understood theoretically,
there can be a clean interpretation of the experimental data, This is of
interest for nuclear structure and nuclear astrophysics
\cite{Bau01,Ver01,Bau02}. It is the purpose 
of this lecture to give an overview of the theoretical methods and to discuss
the experimental results.

Multiple electromagnetic excitation can also be important. We mention two
aspects: it is a means to excite new nuclear states, like the double
phonon giant dipole resonance \cite{Bau02}; 
but it can also be a correction to the
one-photon excitation \cite{Typ01,Typ02,Typ03}. 

The cross sections of such processes rise logarithmically with energy.
This leads to strong effects in the scattering of relativistic beams,
especially heavy ions. Even though this would merit a lecture of its own, 
the production of antihydrogen ($\bar{H^{o}}$) at LEAR and FERMILAB
and electromagnetic processes at the relativistic heavy ion colliders
RHIC and LHC are briefly mentioned.

\section{Theory of electromagnetic excitation and 
\protect\newline dissociation}

\subsection{Inelastic scattering at high energies: one-photon 
\protect\newline exchange,
semiclassical approach and Glauber theory}

In the equivalent photon approximation the cross section for an
electromagnetic process is written as
\begin{equation}
 \sigma = \int \frac{d\omega}{\omega} \: n(\omega) \sigma_{\gamma}(\omega)
\end{equation}
where $\sigma_{\gamma}(\omega)$ denotes the corresponding cross section
for the photo-induced process and $n(\omega)$ is called the equivalent
photon number. For high enough beam energies it can be well approximated
by
\begin{equation}
 n(\omega) = \frac{2}{\pi} Z^{2} \alpha \ln \frac{\gamma v}{\omega R}
\end{equation}
where $R$ denotes some cut-off radius. More refined expressions, which
take the dependence on multipolarity, beam velocity or
Coulomb-deflection into account, are available in the literature
\cite{Ber01,Typ02,Win01}.  

The theory of electromagnetic excitation is well developed, for
nonrelativistic as well as relativistic projectile velocities.
In the latter case, an analytical result for all multipolarities was
obtained in ref.~\cite{Win01}. The projectile motion was treated classically
in a straight-line approximation. Using Glauber theory, the projectile
motion can be treated quantally \cite{Ber01,Typ03,Ber02,Mue01,Bau03}. 
This gives rise to characteristic
diffraction effects. The main effect is due to the strong absorption
at impact parameters less than the sum of the two nuclear radii.
Effects due to nuclear excitation have also to be taken into account
\cite{Mue01,Hen01}.
They are generally small and show a characteristic angular dependence,
which can be used to separate such effects from electromagnetic excitation.

\subsection{Higher order electromagnetic effects, \protect\newline
small $\xi$-approximation}
\label{2.2}

Higher order effects can be considered in a coupled channels approach,
or using higher order perturbation theory. This involves a sum over
all intermediate states n 
considered to be
important. 
Another approach is to integrate the time-dependent 
Schr\"{o}dinger equation directly for a given model Hamiltonian \cite{Esb01}.

If the collision is sudden one can neglect the time
ordering in the usual perturbation approach. The interaction can be summed up 
to infinite order. In order to obtain the excitation cross section, one has to
calculate the matrix-element of this operator between the initial and final
state (i.e.~the intermediate states n do
not appear explicitly). A related
approach was developed for small values of $\xi$ in
Refs.~\cite{Typ01,Typ02,Typ03}. In a simple zero range model 
for the neutron-core interaction, analytical results were obtained 
for $1^{st}$ and $2^{nd}$ order electromagnetic excitation \cite{Typ01}.

In such a model --- a prototype for a loosely bound system, like the
deuteron --- there is only one bound state (s-wave). The continuum
states are plane waves, except that the $l=0$ partial wave is modified
by the short range potential. Electromagnetic excitation is treated
in the semiclassical straight line approximation. This system is
described by a few scaling variables, the binding energy
\begin{equation}
 E_{0} = \frac{\hbar^{2} \eta^{2}}{2 m} \: , 
\end{equation}
the strength parameter
\begin{equation} \label{chidef}
 \chi = \frac{2 Z_{X} Z_{eff}^{(1)} e^{2}}{v b \hbar k}
 \qquad (a = c + n)
\end{equation}
where
\begin{equation}
 Z_{eff}^{(1)} = -\frac{Z_{c} m_{n}}{m_{c} + m_{n}}
\end{equation}
and
\begin{equation}
 k = \sqrt{\eta^{2} + q^{2}}
\end{equation}
with $q=$~wave number in the continuum. $Z_{X}$ is the charge number 
of the target.
The dipole excitation
amplitude is given by
\begin{equation}
 a_{f0}(\vec{q}) = a_{f0}^{(1)} (\vec{q}) + a_{f0}^{(2)} (\vec{q}) 
 = \sum_{\lambda \mu} C_{\lambda \mu} (q,\eta,\chi, \xi) 
     Y_{\lambda \mu}(\hat{q}) \: .
\end{equation}
The coefficients $C_{\lambda \mu}$ are given analytically (eqs.~27 -- 32,
Ref.~\cite{Typ01}). Higher order effects depend essentially on the
strength parameter $\chi$. From eq.~\ref{chidef} it is seen that
higher order effects become more important with $Z_{X}$, 
decreasing impact parameter $b$ and velocity $v$ of the projectile,
as is intuitively obvious.

\section{Discussion of experimental results for nuclear 
\protect\newline structure}

In recent years, electromagnetic excitation of intermediate energy (exotic)
beams has been developed into a useful spectroscopic tool \cite{Mot01,Sch01}. 
By measuring
the excitation energies of the first $2^{+}$ states and the corresponding
$B(E2)$-values, nuclear structure effects, like deformation, can be
studied in a unique way for nuclei far off stability.
Electromagnetic excitation of the $1^{st}$ excited state in ${}^{11}$Be
has been studied experimentally at GANIL \cite{Ann01}, RIKEN \cite{Nak01}
and MSU \cite{Fau01}. This is a good test
case, since the $B(E1)$-value of the corresponding ground-state transition
is known already. Theoretical considerations \cite{Ber03,Typ04}
show that higher order effects are expected to be small.

Coulomb dissociation of exotic nuclei is a valuable tool to determine
electromagnetic matrix-elements between the ground state and the nuclear
continuum. The excitation energy spectrum of the ${}^{10}$Be+n system
in the Coulomb dissociation of the one-neutron halo nucleus ${}^{11}$Be
on a Pb target at $72\cdot$A~MeV was measured \cite{Nak02}. 
Low lying E1-strength
was found. In a similar way, the Coulomb dissociation of the 2n-halo
nucleus ${}^{11}$Li was studied in various laboratories. In an
experiment at MSU \cite{Iek01}
the correlations of the outgoing neutrons were studied.
Within the limits of experimental accuracy, no correlations were found.

\section{Coulomb dissociation as a tool to measure radiative capture
processes relevant to nuclear astrophysics}

In nuclear astrophysics, radiative capture reactions of the type
\begin{equation}
 b + c \to a + \gamma
\end{equation} 
play a very important role. They can also be studied in the time-reversed
reaction
\begin{equation}
 \gamma + a \to b + c \: ,
\end{equation}
at least in those cases where the nucleus $a$ is in the ground state.
As a photon beam, we use the equivalent photon spectrum which is provided
in the fast peripheral collision. Recent reviews, both from an experimental
as well as theoretical point of view have been given \cite{Bau01}, so we want
to concentrate here on a few points.

The ${}^{6}$Li $\to$ $\alpha$ + d Coulomb dissociation has been a test
case of the method, as reviewed in Ref.~\cite{Bau01}. 
The issue of primordial ${}^{6}$Li as a test of big bang nucleosynthesis
has recently been discussed in Ref.~\cite{Nol01}. The abundances of the
lighter isotopes, notably ${}^{4}$He but also ${}^{2}$H, ${}^{3}$He, and
${}^{7}$Li have all been found to be consistent with the primordial level
predicted by the big bang nucleosynthesis over a fairly narrow range of 
the baryon to photon ratio $\eta$: $2.5\cdot 10^{-10} < \eta <
6\cdot 10^{-10}$. ${}^{6}$Li has the next highest predicted primordial
abundance. The observation of primordial ${}^{6}$Li is not within reach of
present experimental capabilities, but must be subjected to future
techniques \cite{Nol01}. The key nuclear input is the cross section for the
capture reaction $\alpha$(d,$\gamma$)${}^{6}$Li at $E \approx 60 - 400$~keV.
The astrophysical S-factor 
has been studied theoretically in 
Ref.~\cite{Muk01} and \cite{Typ05}. In Ref.~\cite{Muk01} a significant
energy dependence of $S_{24}$ at astrophysical energies was found.
At energies less than 110~keV, the E1 component was found to dominate over
the E2 component. The agreement with the experimental results from
Coulomb dissociation \cite{Kie02} is quite good.

The ${}^{7}$Be(p,$\gamma$)${}^{8}$B radiative capture reaction is relevant
to the solar neutrino problem. It determines the production of ${}^{8}$B
which leads to the emission of high energy neutrinos. There are direct
capture measurements, for a recent one see \cite{Bog01}. 
A significantly smaller value
than previously adopted has been found there. Coulomb dissociation
of ${}^{8}$B $\to$ ${}^{7}$Be + p has been studied at intermediate energies
at RIKEN \cite{Mot02}, MSU \cite{Kel01} and GSI \cite{Sue01}, 
and, at rather low energies, at Notre Dame \cite{vSc01}.
Up to now, only the RIKEN experiment has extracted an astrophysical
S-factor from the data. We note that a model-independent separation of E1
and E2 components is possible by a careful study of angular distributions
and correlations \cite{Kel01,Gai02,Mot03}. 
Furthermore, at RIKEN or higher energies, nuclear or
higher order electromagnetic effects are virtually negligible. It can be
hoped that a consistent picture will emerge for all these studies under
different experimental conditions leading to the same astrophysical
S-factor.

Other interesting cases could be the ${}^{7}$Li $\to$ $\alpha$ + t
and ${}^{7}$Be $\to$ $\alpha$ + ${}^{3}$He Coulomb dissociation reactions.
New theoretical results on the radiative capture reactions are given
in \cite{Iga01}.

Nucleosynthesis beyond the iron peak proceeds mainly by the r- and s-processes
(rapid and slow neutron capture) \cite{Rol01,Cow01}. 
To establish the quantitative details
of these processes, accurate energy-averaged neutron-capture cross
sections are needed. Such data provide information on the mechanism
of the neutron-capture process and time scales, as well as temperatures
involved in the process. The data should also shed light on neutron
sources, required neutron fluxes and possible sites of the
processes (see Ref.~\cite{Rol01}).

With the new radioactive beam facilities (either fragment separator or
ISOL-type facilities) some of the nuclei far off the valley of stability,
which are relevant for the r-process, can be produced. In order to assess
the r-process path, it is important to know the nuclear properties like
$\beta$-decay half-lifes and neutron binding energies. Sometimes, the
waiting point approximation \cite{Rol01,Cow01}
is introduced, which assumes an (n,$\gamma$)-
and ($\gamma$,n)-equilibrium in an isotopic chain. It is generally believed
that the waiting point approximation should be replaced by dynamic
r-process flow calculations, taking into account (n,$\gamma$), ($\gamma$,n)
and $\beta$-decay rates as well as time-varying temperature and neutron
density. In slow freeze-out scenarios, the knowledge of (n,$\gamma$)
cross sections is important.

In such a situation, the Coulomb dissociation can be a very useful tool
to obtain information on (n,$\gamma$)-reaction cross sections on
unstable nuclei, where direct measurements cannot be done. Of course,
one cannot and need not study the capture cross section on all the nuclei
involved; there will be some key reactions of nuclei close to
magic numbers. Quite recently, it was proposed \cite{Gai01} to use the Coulomb
dissociation method to obtain information about (n,$\gamma$) reaction
cross sections, using nuclei like ${}^{124}$Mo, ${}^{126}$Ru, ${}^{128}$Pd
and ${}^{130}$Cd as projectiles. The optimum choice of beam energy will
depend on the actual neutron binding energy. Since the flux of equivalent
photons has essentially an $\frac{1}{\omega}$ dependence, low neutron
thresholds are favourable for the Coulomb dissociation method. Note
that only information about the (n,$\gamma$) capture reaction to the
ground state is possible with the Coulomb dissociation method. The
situation is reminiscent of the loosely bound neutron-rich light nuclei,
like ${}^{11}$Be and ${}^{11}$Li. In these cases, the Coulomb dissociation
has proved very useful (see the discussion above).

In Ref.~\cite{Typ01}
the $1^{st}$ and $2^{nd}$ order Coulomb excitation amplitudes
are given analytically in a zero range model for the neutron-core
interaction (see section~\ref{2.2}). 
This can be very useful to assess how far one can go down
in beam energy and still obtain meaningful results with the Coulomb
dissociation method. I.e., where the $1^{st}$ order amplitude
can still be extracted experimentally without being too much disturbed
by corrections due to higher orders. 
For future radioactive
beam facilities, like ISOL od SPIRAL, the maximum beam energy is an
important issue. We propose to use the handy formalism of Ref.~\cite{Typ01}
to assess, how far one can go down in beam energy.
For Coulomb dissociation with two charged particles in the final
state, like in the ${}^{8}$B $\to$ ${}^{7}$B + p experiment with
a 26~MeV ${}^{8}$B beam \cite{vSc01} such simple formulae seem to be
unavailable and one should resort to the more involved approaches
mentioned in section \ref{2.2}.

A new field of application of the Coulomb dissociation method can be
two nucleon capture reactions. Evidently, they cannot be studied
in a direct way in the laboratory. Sometimes this is not necessary, where the
relevant information about resonances involved can be obtained by
other means (transfer reactions, etc.), like in the triple $\alpha$-process.

Two-neutron capture reactions in supernovae neutrino bubbles are studied
in \cite{Goe01}. 
In the case of a high neutron abundance, a sequence of two-neutron
capture reactions, ${}^{4}$He(2n,$\gamma$)${}^{6}$He(2n,$\gamma$)${}^{8}$He
can bridge the $A=5$ and 8 gaps. The ${}^{6}$He and ${}^{8}$He nuclei
may be formed preferentially by two-step resonant processes through their
broad $2^{+}$ first excited states \cite{Goe01}. Dedicated Coulomb 
dissociation experiments can be useful. Another key reaction can be the
${}^{4}$He($\alpha$n,$\gamma$) reaction \cite{Goe01}. 
The ${}^{9}$Be($\gamma$,n) reaction
has been studied directly (see Ref.~\cite{Ajz01}) 
and the low energy $s\frac{1}{2}$
resonance is clearly established. Despite this, a ${}^{9}$Be Coulomb
dissociation experiment could be rewarding (cf. also Ref.~\cite{Kal01}).
Other useful information is obtained from (e,e') and (p,p') reactions on
${}^{9}$Be \cite{Kue01}.

In the rp-process, two-proton capture reactions can bridge the waiting 
points \cite{Bar01,Goe02,Sch02}. From the ${}^{15}$O(2p,$\gamma$)${}^{17}$Ne, 
${}^{18}$Ne(2p,$\gamma$)${}^{20}$Mg and ${}^{38}$Ca(2p,$\gamma$)${}^{40}$Ti
reactions considered in Ref.~\cite{Goe02}, 
the latter can act as an efficient reaction
link at conditions typical for X-ray bursts on neutron stars.
A ${}^{40}$Ti $\to$ p + p + ${}^{38}$Ca Coulomb dissociation experiment
should be feasible. The decay with two protons is expected to be
sequential rather than correlated (``${}^{2}$He''-emission).
The relevant resonances are listed in Table~XII
of Ref.~\cite{Goe02}.
In Ref.~\cite{Sch02} it is found that in X-ray bursts 2p-capture reactions
accelerate the reaction flow into the $Z \geq 36$ region considerably.
In Table~1 of Ref.~\cite{Sch02} nuclei on which 2p-capture reactions may occur,
are listed; the final nuclei are ${}^{68}$Se, ${}^{72}$Kr, ${}^{76}$Sr,
${}^{80}$Zr, ${}^{84}$Mo, ${}^{88}$Ru, ${}^{92}$Pd and ${}^{96}$Cd
(see also Fig.~8 of Ref.~\cite{Bar01}). It is proposed to study the Coulomb
dissociation of these nuclei in order to obtain more direct insight
into the 2p-capture process.

\section{Antihydrogen production and electromagnetic 
\protect\newline processes at relativistic heavy ion colliders}

The process
\begin{equation}
 \gamma + \bar{p} \to \bar{H^{o}} + e^{-}
\end{equation}
leads to antihydrogen \cite{Mun01,Bau04}. 
The photons can come from the Coulomb field of a target nucleus, the
antiprotons are available as a medium energy beam. Thus the produced
$\bar{H^{o}}$ will have essentially the same velocity as the incoming
$\bar{p}$ beam. This was realized at LEAR with
antiprotons impinging on a Xe (Z=54) cluster target \cite{Bau05}. 
In this way antihydrogen
atoms were produced and detected for the first time. Presently, $\bar{H^{o}}$
atoms are produced at FERMILAB with essentially the same method. It
remains to be seen whether further experiments (e.g. measuring the
Lamb-shift of $\bar{H^{o}}$ \cite{Mun01}) are possible with these fast neutral $\bar{H^{o}}$ beams.

A more accurate method will be the production of cold antihydrogen in traps.
This will offer the possibility to use LASER spectroscopy and measure the
spectrum of $\bar{H^{o}}$ with very high precision (e.g. the 1s-2s
transition). This can be a significant test of the CPT-theorem.

A related process might be important
in the future relativistic heavy ion colliders
like RHIC at Brookhaven National Laboratory and LHC at CERN. The capture
of electrons
\begin{equation}
 Z + Z \to (Z + e^{-}) + e^{+} + Z
\end{equation}
leads to a change of the charge state of the circulating naked ions and
therefore to a beam loss and luminosity decrease. The cross section for
this process scales approximately with $Z^{7}$ and will be of the order of
100~b for heavy systems (like Au-Au (Z=79) or Pb-Pb (Z=82)).
Another source of beam loss with cross sections of a similar order of
magnitude is the excitation of the giant dipole resonance, with 
subsequent nucleon emission. The collision of equivalent photons can
be used to study photon-photon collisions (``double Primakoff effect'').
Up to now, this has mainly been done at $e^{+}e^{-}$ colliders.  Due
to the $Z^{4}$ factor in the cross section heavy ion colliders can be
favourable. At RHIC one can explore the invariant mass range up to
several GeV, and at LHC about 100~GeV. Unlike for $e^{+}e^{-}$
colliders there is a strong interaction background, which has to be
taken care of. This is studied in the LoI for the proposed FELIX
detector at LHC \cite{FEL01}. 
Such $\gamma \gamma$ collisions lead to the production
of lepton pairs ($l=e,\mu,\tau$), $C=+1$ mesons ($\pi^{0}, \eta, \eta_{c},
\eta_{b}, \dots$), vector meson pairs, Higgs-bosons, etc. 
This subject would actually be another lecture and we wish to limit ourselves
to these remarks. Further references can be found in \cite{FEL01,Hen02}.

\section{Conclusions}

Peripheral collision of medium and high energy nuclei (stable or
radioactive) passing each other at distances beyond nuclear contact
and thus dominated by electromagnetic interactions are important tools
of nuclear physics research. The intense source of quasi-real
(or equivalent) photons has opened a wide horizon of related problems
and new experimental possibilities to investigate efficiently
photo-interactions with nuclei (single- and multiphoton excitations
and electromagnetic dissociation).

\subsection*{Acknowledgements}

It is a pleasure to thank many colleagues for interesting discussions,
we are particularly grateful to P.~Aguer, C.~A.~Bertulani, M.~Gai, J.~Kiener,
K.~Langanke, T.~Motobayashi, H.~Rebel, K.~S\"{u}mmerer,
F.-K.~Thielemann, D.~Trautmann and M.~Wiescher.

\end{document}